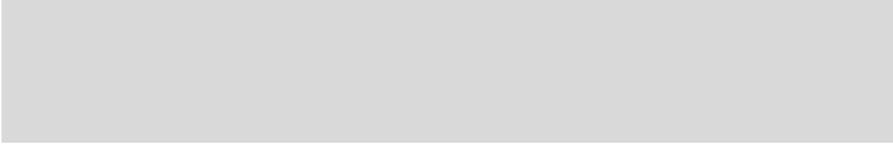

# Analyzing the concept of technical debt in the context of agile software development: A systematic literature review

*Woubshet Nema Behutiye a, Pilar Rodriguez a, Markku Oivo a, Ayşe Tosun b*

"a University of Oulu, Oulu, Finland"
"b Faculty of Computer Engineering and Informatics, Istanbul Technical University, Turkey"

Keywords:
Technical debt
Agile software development
Technical debt management
Systematic literature review

A B S T R A C T

**Context**: Technical debt (TD) is a metaphor that is used to communicate the consequences of poor software development practices to non-technical stakeholders. In recent years, it has gained significant attention in agile software development (ASD). **Objective**: The purpose of this study is to analyze and synthesize the state of the art of TD, and its causes, consequences, and management strategies in the context of ASD. **Research Method**: Using a systematic literature review (SLR), 38 primary studies, out of 346 studies, were identified and analyzed. **Results**: We found five research areas of interest related to the literature of TD in ASD. Among those areas, "managing TD in ASD" received the highest attention, followed by "architecture in ASD and its relationship with TD". In addition, eight categories regarding the causes and five categories regarding the consequences of incurring TD in ASD were identified. "Focus on quick delivery" and "architectural and design issues" were the most popular causes of incurring TD in ASD. "Reduced productivity", "system degradation" and "increased maintenance cost" were identified as significant consequences of incurring TD in ASD. Additionally, we found 12 strategies for managing TD in the context of ASD, out of which "refactoring" and "enhancing the visibility of TD" were the most significant. **Conclusion**: The results of this study provide a structured synthesis of TD and its management in the context of ASD as well as potential research areas for further investigation.

## 1. Introduction

The technical debt (TD) concept was first introduced by Ward Cunningham in 1992 [1] as a means to communicate the challenges arising from the consequences of poorly developed software to non-technical stakeholders. Since then, it has evolved to embrace different aspects of software development that range from architectural and design problems to documentation, people, and deployment issues [2].

Recently, TD has become a popular concept in agile software development (ASD) ([P36], [P21], [3], and [P34]) owing to its economic implications and the specific characteristics of ASD that make it prone to incurring TD. In ASD, the emphasis on delivering functionality quickly often lessens the focus on aspects such as design, good programming practices, or test coverage with consequences leading to the rise of TD, which then needs to be addressed in later phases of the development process [P22]. For example, deviations in code and design made in favor of immediate delivery may incur TD in the long run ([P27], [P4], and [P1]). As a result, companies that apply ASD may face increased release durations, growing costs, and a failure to maintain quality. Moreover, in some cases, the economic implications can go as far as a market loss due to hampering relationships with business stakeholders [P20]. However, when properly managed and controlled, TD can be utilized as an agile strategy to gain business opportunities and be paid back later ([P5] and [P21]). Hence, understanding TD and its implications is important in ASD.

Several studies have been conducted to understand the different perspectives and implications of the TD metaphor in ASD (e.g. [P18], [P22], [P27], and [P21]). However, their findings suggest that a solidified understanding of the concept of TD and its management in ASD is still lacking. The agile community lacks rigorously evaluated guidelines for characterizing, managing and prioritizing TD [P22]. Furthermore, the literature on the area has not been well structured to present a clarified image of TD and its management within the context of ASD. Despite the existence of secondary studies investigating TD and its management in software development (Tom et al. [4], [5]; Li, et al. [6]; Alves et al. [7]; Ampatzoglou et al. [8]), none of the studies focused their analysis within the context of ASD. This makes it difficult for agile practitioners to identify causes, consequences and TD management (TDM) strategies specific to ASD.

For instance, specific causes of TD in ASD, which are not revealed by the cited secondary studies, include overlooked estimates of sprints [P28], parallel development in isolation [P5] and organizational gaps among business, operational and technical stakeholders [P22]. In such cases, general TDM strategies found in earlier secondary studies such as code and dependency analysis, and cost-benefit analysis ([6], [7], and [8]) may not be helpful enough. However, strategies specifically designed to manage TD in ASD, such as responsibility driven architecture [P7], acceptance test reviews [P12], and defining a common 'Definition of Done (DoD)' [P14], are pointed to be helpful in the context of ASD. Our work complements earlier secondary studies by identifying new primary studies (16), which are not included in previous secondary studies, and analyzing primary studies differently in order to focus on particularities of ASD. We reveal a new insight into unique causes of TD and TDM strategies specific to ASD such as 'the common DoD' [P14], the MAKEFLEXI [P17], and propagation cost model [P8]. Additionally, the fact that "*agile development appears to be more prone to technical debt accumulation compared to traditional software development approaches, due to its delivery-oriented focus*" [9] also raises the significance of the topic.

Our study identifies and structures the body of knowledge of TD in the context of ASD by aggregating scientific contributions in a systematic way and determining research gaps for future research activities ([10], [23]). The objectives of this study are as follows:

- Analyzing the scientific literature of TD in the context of ASD in order to:
  - Identify, classify, and structure the scientific studies of TD in the context of ASD.
  - Investigate the research type and the kind of contributions of the research of TD in ASD and assessing quality of these studies.
  - Identify facets that are reflected in the TD metaphor in the context of ASD.
  - Identify, analyze, and aggregate the results of the different areas of research that investigate TD in the context of ASD.
- Identifying and analyzing the causes and consequences of TD in the context of ASD.
- Identifying TDM strategies applied to control TD in the context of ASD.
- Exploring potential research gaps within the area in order to guide future research initiatives.

The contribution of our work is twofold. On the one hand, our study provides researchers with a structured understanding of the state of the art of TD within the context of ASD upon which they can base their studies on the topic, as well as recognizing research gaps that require further study. On the other hand, the findings from this systematic literature review (SLR) will help practitioners clarify their understanding of the concept of TD and its causes and consequences as well as the management strategies applied to control it within ASD.

The rest of the paper is organized as follows: Section 2 presents background and related works on TD and ASD. Section 3 describes the research methodology, including a discussion of threats to validity and the countermeasures taken to minimize their effects. Section 4 presents the results of the study, and Section 5 discusses the results in light of related work. Finally, the conclusions of the study are presented in section 6.

## 2. Background and related work

This section first describes the core ideas of TD and ASD and analyzes the reasons why TD is especially relevant in the context of ASD. Then, previous reviews of the TD and ASD literature are summarized and the need that motivated this study is justified.

### *2.1. Background*

*2.1.1 Technical debt (TD)*

Cunningham (1992) [1] introduced the notion of TD by drawing an analogy to financial debt: "*Shipping first-time code is like going into debt. A little debt speeds development so long as it is paid back promptly with a rewrite. Objects make the cost of this transaction tolerable. The danger occurs when the debt is not repaid. Every minute spent on not-quite-right code counts as interest on that debt.*" Since then, a wide range of definitions have been suggested by different authors with the aim of clarifying the TD concept ([P27], [4], [11], [P4], [36]). For instance, McConnell (2007) [11] defined TD as "*a design or construction approach that's expedient in the short term but that creates a technical context in which the same work will cost more to do later than it would cost to do now (including increased cost over time).*" Others extend the definition to describe different manifestations of TD. For example, Hossein and Ruhe (2015) [36] define requirements debt, which is the manifestation of TD in requirements, as: "*the trade-offs in requirements specification that are consequences of the intentional strategic decisions for immediate gains or unintentional changes in the context that have an impact on the future cost of the project.*" Recently, the definition of TD was refined as, "*in software-intensive systems, technical debt consists of design or implementation constructs that are expedient in the short term, but set up a technical context that can make a future change more costly or impossible. Technical debt is a contingent liability whose impact is limited to internal system qualities, primarily maintainability and* evolvability" in Dagstuhl seminar of managing TD [39].

Among the various manifestations of TD, code decay and architectural deterioration are the most recognized dimensions in the literature [5]. However, TD can be related to other activities as well, including design, documentation, and testing [12]. For example, in a recent study, Alzaghoul and Bahsoon (2013) [13] introduce a new perspective of TD in cloud computing services related to the selection, composition, and operation of services. Non-compliance in service level agreements (SLA), quick selection decisions with a lack of consideration for risk reduction and probable future changes, and under-utilization of web service capacity were identified as TD dimensions in the context of cloud services [13].

Understanding TD from both the theoretical and practical perspectives is important in advancing the state of the art of this concept [3]. Owing to the broad perspectives reflected in the concept, practitioners and researchers in academia have proposed various classifications of TD to help with the understanding and management of it. Among these, Steve McConnell's taxonomic classification of TD [11] and Martin Fowler's TD quadrants [35] are notable [14]. McConnell (2007) [11] classifies TD as "intentional" and "non-intentional" on the basis of the reasons for the accrual of TD. In this classification, non-strategic TD that results from the developer's coding inexperience or errors in design approaches are categorized as unintentional TD. In contrast, when TD is incurred strategically to facilitate releases, it is categorized as intentional TD.

Fowler's TD quadrant [35] classifies TD as "prudent" and "reckless" debts that can each be incurred in either an inadvertent or a deliberate way. Prudent debt represents TD that is incurred proactively in order to achieve quick releases. In this case, when the teams have plans to deal with the consequences, the debt is incurred deliberately. However, Fowler [35] also argues that TD can be manifested as both prudent and inadvertent at the same time. In such cases, the lack of knowledge is mainly attributed to TD. For instance, developers may come to realize that a different design approach should have been taken at some point in the development process. Reckless TD, on the other hand, highlights debts that are either acquired deliberately as the result of a shortage of time (and which development teams have no intention of solving in the future) or incurred unintentionally as "inadvertent" reckless debt.

*2.1.2 Agile software development*

Agile methods were initially born to overcome the challenges that were faced by traditional, plan-driven, heavyweight software development methods in responding to ever-changing business demands ([15], [16], [17], and [18]). The focus of the traditional approaches on eliminating changes rather than embracing them was a challenge when it was necessary to address a growth in customers' expectations and market demands [17]. Traditional methods such as the waterfall model emphasize fixed and predetermined requirements where the development is process-centric. As a result, it was not possible for customers to give frequent feedback, making it hard to clarify misunderstandings and the needs of change in the development process [19]. Although ASD caused some initial controversies, it has become more and more popular among practitioners [37]. Abrahamson et al. (2002) [20] describe ASD as an incremental, straightforward (easy to learn) development, which is characterized by an adaptive nature that is easy to modify and by cooperation and open communication between developers and customers. In general, ASD entails a specific group of software development methods that is iterative and incremental and driven by a set of values established through the Agile Manifesto [16], which is focused on embracing change and people's collaborations in software development. Some of the agile methods used in the industry include eXtreme Programming (XP), Scrum, feature-driven development, and crystal methods [20].

One of the key principles of ASD is delivering working software more frequently. Although the Agile Manifesto also highlights "*continuous attention to technical excellence and good design enhances agility*" in one of its 12 principles, it has been found that delivering software as fast as possible without adequate attention to the engineering practices can present the challenge of accumulating TD in the context of ASD [P38].

*2.1.3 Understanding TD in agile software development (ASD)*

Understanding TD in ASD is essential to make appropriate TDM decisions at the right time. For example, in cases where there is no agreement as to the type and amount of an acceptable level of TD, agile developers can make assumptions and take paths that negatively affect the development. Consequently, the productivity of the team may decrease, the quality of the software product may degrade, rework costs may increase, and, at times, business relationships might also suffer due to all of these [21]. However, when ASD teams are aware of TD and its implications, they can instead use the concept to their advantage. In such cases, agile teams can plan in advance how to deal with TD and use it to gain business opportunities [2].

Different reasons have been proposed to explain why ASD is prone to TD. For example, ASD puts less emphasis on documentation practices and instead prioritizes the delivery of working software, which makes it prone to TD [9]. Additionally, insufficient attention given to software architecture makes ASD prone to TD [P25]. In ASD, the drive for efficiency and expediency often leads to less focus on architectural models [P35]. Likewise, the findings from a recent mapping study on continuous deployment (a concept closely related to ASD) reveal that, as a consequence of trade-offs between the fast deployment of software and poor development, testing, and quality assurance practices, organizations tend to acquire TD over time [25].

Therefore, agile practitioners need to be particularly aware of the constituents of TD, its economic importance, the strategies that can be applied to paying it down when using ASD, and its other related implications in ASD. Analyzing TD in a broad domain such as ASD requires a thorough investigation of the relevant studies reporting on the various aspects, including a formalization of the concept and TDM strategies and experiences. One way to do this is to structure and synthesize the existing knowledge on the topic and assess the research gaps. The arguments that ASD is prone to TD ([9], [25]), the economic and technical implications of TD in ASD, and the absence of secondary studies investigating TD in the context of ASD motivated us to conduct this SLR.

*2.2. Related work*

As per our knowledge, there are no secondary studies investigating the state of the art of TD in the context of ASD prior to our study. However, different secondary studies have been conducted on TD and its management (Tom et al. (2012) [4]; Tom et al. (2013) [5]; Li et al. (2015) [6]; Alves et al. (2015) [7]; Ampatzoglou et al. (2015) [8]) and on agile development (Dybå et al. (2008) [18]).

Tom et al. (2012) [4] conducted an SLR to get a consolidated understanding of TD and determine the positive and negative outcomes associated with it. They identify that code decay and architectural deteriorations are recognized as the major constituents of TD in the academic literature. The authors also propose a theoretical framework that helps uncover elements of TD, establishes boundaries, and identifies the causes behind accruing TD. Budget and resource constraints were cited as potential causes of accumulating TD, and negative consequences in scheduling, risk, and quality were found as the associated outcomes.

As a continuation of the preliminary SLR study [4], Tom et al. (2013) [5] conducted an exploratory case study using a multivocal literature review, supplemented with interviews of practitioners and academics, to gain a more comprehensive understanding of TD and its implications in software engineering. (multivocal literature reviews are comprised of all accessible writings, including non-academic writings on a topic [22].) In this study, the authors develop a taxonomical hierarchy of TD and a theoretical framework to visualize dimensions, attributes, precedents, and outcomes of TD and further clarify the boundaries of the concept.

Li et al. (2015) [6] include peer-reviewed studies published between 1992 and 2013 in their systematic mapping (SM) study to get a holistic understanding of TD and its management. The authors identify 10 types of TD, of which code debt is the most studied type. Additionally, they identify and classify five categories of TD notions, namely, metaphor, property, cause, effect, and uncertainty. The metaphor idea relies on metaphors adapted from the field of economics to describe TD; the properties idea describes characteristics of TD; the cause idea explains the related reasons for incurring TD; the effect idea describes the related effects of incurring TD; and the uncertainty idea reflects the ambiguous nature of TD. Regarding compromised quality attributes, the authors find that most studies argue that TD negatively affects maintainability.

Ampatzoglou et al. (2015) [8] conducted an SLR in order to examine the state of the art on TDM, with a focus on the financial aspects of TD in software engineering. The authors find that "principal" and "interest" were the most popular financial terms used with TD. "Principal" describes the effort required to resolve TD, whereas "interest" describes the additional cost required to pay back incurred TD in later phases. Moreover, they identify real options, portfolio management, cost/benefit analysis, and value-based analysis as financial strategies for managing TD.

Recently, Alves et al. (2015) [7] conducted an SM study to investigate the types of TD, as well as the TD identification and management strategies, which have been proposed in the literature. The authors identify various indicators of TD, such as god class and duplicate code, which support the identification of specific types of TD. However, software visualization techniques were the least used to identify TD. Regarding TDM strategies, the authors identify the portfolio approach and cost-benefit analysis as the most frequently cited strategies.

Dybå and Dingsøyr (2008) [18] conducted a systematic review of ASD publications in order to investigate the empirical findings of ASD studies, provide an overview of studied topics, evaluate the strength of the findings, and articulate the implications for academia and practitioners. The authors identify four aspects of ASD: introduction and adoption, human and social factors, comparative studies of ASD, and the perception of agile methods among developers, customers, and students. However, despite the fact that the TD concept has existed since 1992 [1] and the significance and popularity of the concept in ASD [P18], it is not reported in the paper. Perhaps this can be attributed to the fact that research interests in TD have only begun receiving more attention in recent years.

Our work differs from these secondary studies in terms of the research goal perspective, and, in some cases, the research method employed. Unlike all the aforementioned secondary studies, our study focuses on understanding and structuring the literature on TD within the specific context of ASD. Despite the existence of strategies specifically designed to manage TD in the context of ASD (e.g. common DoD), and specific ASD processes and practices that could be prone to incurring TD (e.g. inaccurate estimations when planning the sprint, incomplete user stories), the particularities of TD in ASD have not been the object of study of previous secondary studies. We base our analysis on peer-reviewed articles to identify important research areas of interest and determine the causes, consequences, and management strategies of TD in the context of ASD. Our work complements the existing secondary studies by introducing a new perspective of TD and its management in ASD. In comparison with the research methods applied in the reviews discussed previously, there are also some differences. For instance, Tom et al. (2013) [5] employed multivocal literature reviews and included blogs and publications prior to 2011. Other studies, such as [6] and [7], employed SM as the main research method. However, our work employs an SLR and is limited to peer-reviewed scientific publications between 1992 and June 2014.

## 3. Research method

We used the SLR guidelines produced by Kitchenham and Charters (2007) [23] to conduct this study. Additionally, the SM guideline by Peterson et al. (2008) [10] was also taken into consideration while carrying out the keywording process in order to identify the research areas investigating TD in the context of ASD. The steps followed in the study are described in the following subsections.

### 3.1. Definition of research questions

The objective of this study is to identify, classify, and analyze the state of the art of TD in the context of ASD. Accordingly, four research questions were formulated, as shown in Table 1.

The purpose of the first research question is to analyze the state of the art of TD in ASD from different aspects. In addition to bibliographic information and research characteristics (such as research type, contribution type, scientific rigor, and industrial relevance), we aim to identify definitions that refer to TD in the context of ASD as well as research areas that focus on the topic. The second research question aims to identify the causes and consequences of incurring TD in ASD. The third research question intends to identify TDM strategies in ASD in terms of practices, tools, frameworks, and other contributions. Finally, with the fourth research question, we aim to identify research areas that require further study.

**Table 1. Research questions**

| Research question | Aim |
|---|---|
| RQ1: How is the research of TD characterized in the context of ASD? | Analyzing the research of TD in the context of ASD. |
| RQ1.1. What is the current state of the research pertaining to TD in the context of ASD in terms of research types, research contributions, and the quality of the reported research? | ○ Identifying, classifying, and structuring the scientific studies of TD in the context of AS. ○ Investigating the research types and the kind of contributions of the research on TD in ASD, and assessing the quality of these studies. |
| RQ1.2. How is the concept of TD defined in the context of ASD? | ○ Identifying facets that are reflected in the TD metaphor in the context of ASD. |
| RQ1.3. What research areas are emphasized in the literature that reports studies of TD in the context of ASD? | ○ Identifying research areas associated with TD in the context of ASD. |
| RQ2: What are the related causes and consequences of accruing TD in ASD? | Identifying the causes and consequences of incurring TD in the context of ASD, as reported in the literature. |
| RQ3: What are the strategies proposed in the literature to manage TD in ASD? | Identifying TDM strategies applied to control TD in the context of ASD. |
| RQ4: What are the existing research gaps in the field of TD in ASD? | Identifying potential research gaps in the context of TD and ASD in order to guide future research initiatives. |

### 3.2. Conducting the search

In order to identify the primary studies, we used electronic database searches with a search strategy that addresses the research questions of the study. Search strings were constructed following the population, intervention, comparison, and outcome (PICO) criteria suggested by Kitchenham & Charters [23]. However, we relied on a combination of the population AND intervention groups of search terms; the comparison and outcome facets of PICO were omitted in the search structure because we are not interested in comparing the interventions or limiting the outcome. The population facet represents search terms that make reference to ASD.

Alternative keywords that refer to ASD were combined to form the population based on the search string used by Dybå and Dingsøyr [18] in their SLR on empirical studies of ASD. Similarly, search terms for the intervention facet were constructed using alternative keywords that refer to the TD concept based on the findings of Tom et al. [5]. The population and intervention search terms were joined using the "AND" Boolean operator to build the search string, as shown in the Table 2.

ACM, Google Scholar, IEEE Xplore, ProQuest, Scopus, and the Web of Science databases were used in the search for primary studies. The rationale behind the selection of these databases is the extensive list of articles, journals, and conference proceedings related to software engineering that they provide. We executed tailored search queries in the six databases mentioned above (depending on the syntax requested by each database) based on the search string and keywords shown in Table 2. The full-text search execution in these databases retrieved a total of 346 relevant studies, as shown in Table 3.

**Table 2. Search string**

| Population* | AND | Intervention** |
|---|---|---|
| software AND | | "Technical debt" |
| (agile | | OR "design debt" |
| OR XP | | OR "code debt" |
| OR | | OR "debt metaphor" |
| "extreme programming" | | OR "architectural debt" |
| OR scrum | | OR "environmental debt" |
| OR crystal | | OR "testing debt" |
| OR dsdm | | OR "knowledge debt" |
| OR fdd | | OR |
| OR "feature driven development " | | "technical debt man*" |
| OR lean) | | |

\* Based on the search strings from Dybå and Dingsøyr, [18]
\*\* Based on the findings of Tom et al. [5]

**Table 3. Number of studies retrieved in databases**

| Databases | Number of retrieved studies |
|---|---|
| ACM | 62 |
| Google scholar | 126 |
| IEEE Xplore | 12 |
| ProQuest | 67 |
| Scopus | 76 |
| Web of Science | 3 |
| Total | 346 |

*3.3. Screening of relevant papers*

The screening process includes both establishing inclusion/exclusion criteria to find relevant studies related to the research area and a further selection process to identify and select primary studies. The database selection process was complemented with snowballing in order to minimize the validity threat of missing relevant primary studies.

*3.3.1 Inclusion and exclusion criteria*
The following inclusion and exclusion criteria were laid out to ensure the inclusion of relevant studies that are within the scope of this study and address the research questions.

Inclusion criteria:

- Publications that contribute to the knowledge area of both TD and ASD.
- Peer-reviewed publications.
- Scientific papers (including experience reports).
- Publications that are written in the English language.

Exclusion criteria:

- Topics that are not related to TD and ASD (e.g., if the paper merely mentions the concepts of TD and ASD without an investigation of the topic of TD in the context of ASD).
- Studies published in unrecognized (non-scientific) venues that are different from journal or conference publications, such as prefaces, article summaries, overhead presentations, interviews, short papers, introductions to special issues, tutorials, theses, books, and book chapters.
- Duplicates of the same study (in this case, the most complete version of the study was selected).

In addition, this SLR is limited to scientific studies published between 1992 and June 2014. Our rationale for selecting this period relies on the fact that the TD concept was first coined by Cunningham in 1992. Additionally, software development practices that are the basis of ASD have also been in use long before the official declaration of the Agile Manifesto in 2001. Hence, we aim to include relevant studies of ASD that have been published prior to the official declaration of the Manifesto, if any.

*3.3.2 Study selection*

The study selection process started with the exclusion of duplicates, followed by checking whether the study was a scientific article or not. Following this, retrieved studies were screened based on whether the study contributes to the body of knowledge of TD in ASD. This step was carried out in different rounds depending on the parts of the article that we needed to read in order to make sure that the study was within the scope of our research. Thus, we started selecting studies on the basis of reading titles and keywords. If we were unsure about the nature of the study, we proceeded to reading the abstract, the introduction, and the conclusion; lastly, we consulted the full text of the paper if needed.

In order to increase the reliability of the selection process, a pilot was performed of the inclusion and exclusion criteria on 20 randomly selected studies. Two evaluators (the first and second authors) followed the screening process independently on the 20 randomly selected studies. The consistency of the selection process for each criterion was evaluated through an inter-rater agreement calculation using Fleiss' Kappa [24]. Furthermore, in order to describe the relative strength of agreement of the Kappa statistics, we adapted Landis & Koch (1977) [32]. According to [32], Kappa values < 0 indicate no agreement and values from 0 – 0.20 indicate slight agreement, values from 0.21 – 0.40 show fair agreement, values from 0.41– 0.60 indicate moderate agreement, values from 0.61– 0.80 show substantial agreement, and those from 0.81 to 1 indicate an almost perfect agreement.

Results from the inter-rater agreement showed that there was a moderate agreement (Fleiss' Kappa. K=0.5) in identifying papers as scientific and a perfect agreement (Fleiss' Kappa K= 0.897) for determining the relevance of studies to the research area. We further discussed the differences in evaluation in a meeting to clarify disagreements and reach a common understanding on the selection process.

Once the selection process was validated, it was applied on the 346 papers retrieved from our search. 71 studies were duplicates and 115

documents were non-scientific. Non-scientific papers excluded in the process were in the form of books, theses, tutorials, keynotes, blogs, and conference announcements. Following the exclusion of duplicates and non-scientific papers, the remaining 160 papers were screened in order to check whether they focused on TD in the context of ASD based on title and keywords. At this stage, 69 were excluded, eight primary studies were selected, and 83 studies were passed to the next step. The selection then proceeded by checking the abstract, introduction, and conclusion of each study, and reading the paper fully if need be, as mentioned in the above paragraph. As a result, 31 primary studies were found at the end of the selection process.

*3.3.3 Snowballing process*

The study selection process was complemented with the backward snowballing process, which involves checking the references of the 31 primary studies. We also checked recent works from the main authors of the primary studies, whenever it was mentioned that there was a work in progress, in order to determine whether there were any other relevant studies. As a result, four additional papers were found and included (Dos et al. [P38]; Frank et al. [P14]; Antinyan et al. [P20]; Holvitie et al. [P33]) at this stage. A further investigation into the reasons behind the absence of these papers in the search results indicated that the papers were not indexed in the databases that were used in our search execution. Moreover, it was observed that two of the papers appeared in XP/Agile20XX conferences in different years, while the other two papers were not indexed at the time of our search execution.

The XP conference series, established in 2002, was the first conference dedicated to agile and lean processes in software engineering [38]. Similarly, the Agile20XX conference, organized by the Agile Alliance, is an annual agile conference that has been carried out since 2002. Hence, we decided to do a further manual search of XP/Agile20XX conferences from 2002-2014, to ensure that we did not miss any relevant papers. As a result, we found three additional papers (Birkeland et al. [P15]; Elssamadisy et al. [P28]; and Stolberg, [P6]) from the search extension, supplementing our primary studies. Therefore, a total of 38 primary studies were found at the end of the selection process, including the papers acquired through the snowballing process (see Appendix A). The overall selection process for the primary studies, including snowballing, is shown in Table 4.

*3.4. Data extraction*

In order to answer RQ1, we extracted publication details such as publication year, publication source, research method, research type, contribution type, pertinence, industrial relevance, and research rigor. Additionally, deductive coding was followed to identify definitions of TD in the context of ASD (RQ1.2). Similarly, RQ2 and RQ3 (causes and consequences of TD in ASD and TDM strategies in ASD) were answered by using a deductive approach. An overview of the data extracted to answer the RQs is shown in Table 5, and a further description of the properties is included in Appendix B.

We initially extracted keywords that reflected the contribution of the studies in order to identify topics that are emphasized in the literature on TD and ASD (RQ1.3). The qualitative data and research analysis software NVivo was used to support this process. Coding in NVivo was applied by quoting sentences directly from the primary studies that referred to areas of interest of our study.

In order to validate the data extraction process, the first two authors of the paper carried out a pilot and independently extracted data on five randomly selected primary studies. The findings of the data extraction pilot showed that there were differences in classifying research types and contributions. For instance, the inter-rater agreement calculation using Fleiss' Kappa when classifying the research type as a solution proposal shows that we had a substantial level of agreement (Fleiss' Kappa K= 0.651).

**Table 4. Selection of primary studies**

| Step | Excluded studies | No. of studies left | Selected primary studies | Studies to next stage |
|---|---|---|---|---|
| 1. Search retrieval from databases | 0 | 346 | 0 | 346 |
| 2. Exclusion of duplicates | 71 | 275 | 0 | 275 |
| 3. Exclusion of non-scientific papers | 115 | 160 | 0 | 160 |
| 4. Inclusion based on title and keywords | 69 | 91 | 8 | 83 |
| 5. Inclusion based on abstract | 28 | 55 | 5 | 50 |
| 6. Inclusion based on introduction and conclusion | 4 | 46 | 8 | 38 |
| 7. Inclusion based on reading the paper fully | 28 | 10 | 10 | 0 |
| 8. Snowballing |  |  | 7 |  |
| Total |  |  | 38 |  |

**Table 5. Overview of extracted data**

| Property ID | Property |
|---|---|
| Pr1 | Publication ID |
| Pr2 | Publication source (conference/journal) |
| Pr3 | General type of paper (empirical/theoretical/both) and research method (case study, survey, experiment, etc.) |
| Pr4 | Research type (evaluation, solution proposal, experience, philosophical, and opinion) |
| Pr5 | Pertinence (full, partial and medium) |
| Pr6 | Contribution (advice/implications, framework, guideline, lessons learned, model, theory, and tool) |
| Pr7 | Research rigor (value from 0-3) |
| Pr8 | Industrial relevance (value from 0-4) |
| Pr9 | TD definitions in ASD |
| Pr10 | TD causes in ASD |
| Pr11 | TD consequences in ASD |
| Pr12 | TDM strategies in ASD |

*3.5. Keywording*

Keywording was applied in order to identify and structure the research areas emphasized in studies of TD in the context of ASD (RQ1.3). Keywording is a process that clusters keywords that reflect topics of interest in the primary studies. According to Peterson et al. [10], keywording facilitates the creation of the classification scheme and helps ensure that all aspects of the studies are reflected in the mapping. The keywording followed in this study is presented in Fig. 1. First, the abstracts, introductions, and conclusions of the primary studies were read in order to identify keywords that reflect the areas of interest of the primary studies from the perspective of TD in ASD. Then (step 2), the keywords were clustered into groups to identify recurring patterns. The outcome of this step was an initial insight into topics characterizing the research area. In the third step, we utilized the word frequency query feature of NVivo in order to identify the most frequently occurring keywords. We carried out this step to reconfirm that we were not missing important topics. Following these steps, the classification schema was built up by fully reading the primary studies and taking into account the most frequently used keywords that suggested significant research areas (outcome of steps 2 and 3). As a result, we found five research areas of interest (see Section 4.3).

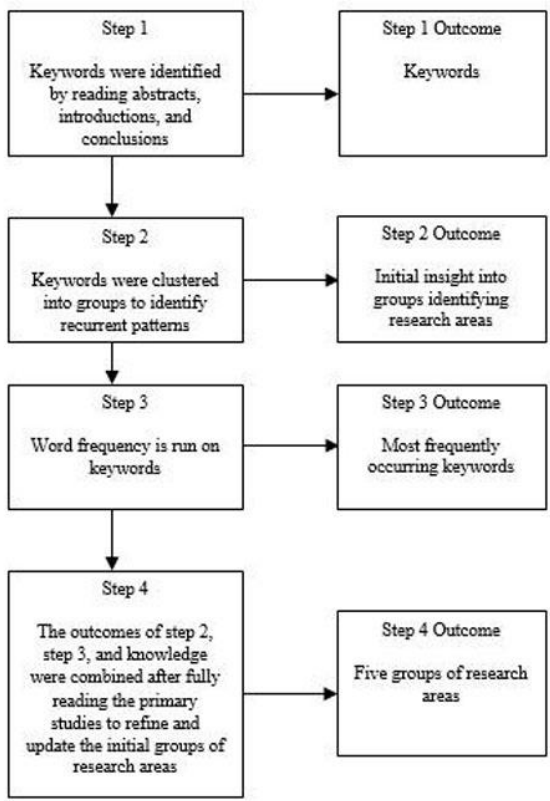

**Fig. 1. Keywording**

### 3.6. Quality assessment

In order to assess the quality of the primary studies, we used the model proposed by Ivarsson and Gorschek (2011) [27] to test research rigor and industrial relevance. The distinction between research rigor and industrial relevance of the model adds an important angle of analysis (e.g., a highly relevant study can have a very low rigor or vice versa [28]), so it helps guide the interpretation of the findings. We evaluated the research rigor of each primary study as the sum of context, study design, and validity values, on a range scale of 0 to 3. Similarly, the industrial relevance of a study was evaluated as the sum of context, research method, scalability, and subject values. The range scale for the total relevance runs from 0 to 4. A further description of the research rigor and industrial relevance aspects is included in Appendix B.

Similar to the data extraction, two of the authors carried out a pilot of the research rigor and industrial relevance assessment criteria on five randomly selected primary studies in order to increase the reliability of the quality assessment. During the pilot, there were differences when evaluating rigor, specifically context, and validity dimensions. While the differences were mainly subjective, a common understanding and clarification on the model was achieved to guide the rest of the quality assessment.

### 3.7. Data synthesis

The data extracted from the primary studies was synthesized by applying thematic synthesis as is suggested by Cruzes & Dybå (2011) [29]. Thematic synthesis applies coding on primary studies; related concepts and findings are labeled and further translated into themes by drawing recurrent patterns. These themes are subjected to further analysis wherein higher ordered themes are created by exploring the relationships among subthemes. We followed a deductive approach based on thematic synthesis in NVivo to extract TDM strategies, the causes and consequences of incurring TD in ASD, and TD definitions. These were further labeled and clustered through conceptual links and relationships to form initial categories. Recurrent patterns were categorized into the initial themes of definitions, causes, consequences, and TDM strategies. These themes were further analyzed and refined to obtain higher-level themes through an inductive thematic synthesis approach, forming categories.

### 3.8. Threats to validity

When considering the validity concerns in our study, the main limitations come from biases in the identification of relevant studies, primary study selection, data extraction, and quality assessment processes. However, in order to ameliorate these threats and increase the reliability of the study, we took the following mitigation actions.

There is an ample amount of non-scientific literature, such as blogs, tutorials, white papers, etc. that discuss TD within the context of ASD. Consequently, the inclusion of such studies might have different results. In order to increase the reliability of the findings of the study and minimize limitations that could come from the inclusion of non-scientific literature, we only included peer-reviewed scientific articles.

In order to mitigate limitations in the identification of primary studies, we applied a systematic search strategy to increase the retrieval of relevant studies in our search [23]. We constructed the search strings based on the research questions and relevant SLRs on TD and ASD. We included all publications between 1992 and June 2014 to increase retrieval of all relevant studies in the research area. We also performed snowballing and applied an extension search through Agile (agile20XX) and XP conferences in order to minimize the chance of missing relevant publications. While the aforementioned measures minimize the risk of omission of relevant papers, we still cannot rule out the possibility that we missed relevant studies (e.g. if there are interventions that are not described as debt and were not hit during the snowballing process).

The potential threat due to the primary study selection process was mitigated by two of the authors carrying out pilots on a total of 20 studies. This was done to build a common understanding of the selection process, clarify differences in the inclusion/exclusion criteria, and improve the reliability of the selection process. Similarly, potential threats from the data extraction and quality assessment processes were mitigated by conducting pilot runs. Two authors conducted the pilot data extraction to mitigate risks that may result in subjective interpretation of data synthesis. A pilot of quality assessment was also performed on five primary studies to minimize the risk of misinterpretation while assessing the research rigor and industrial relevance of primary studies. We believe that these mitigations increase the reliability of the study findings and the conclusions derived from the analysis.

Another potential threat comes from the keywording process used to structure the research areas discussing TD in ASD. It is possible that the results of classification may have overlooked some research areas of interest. However, we undertook a detailed, progressive analysis and refinement process that is complemented with automatic word query findings in the keywording to minimize the risk of misinterpretation of the result and classifications.

## 4. Results and analysis

This section presents the findings from the analysis of the 38 primary studies (with the exception of rigor and relevance assessment that is carried out only on 28 primary studies based on empirical data). In Section 4.1 we provide an overview of the research on TD in ASD (RQ1.1), and in Section 4.2, we identify different aspects reflected in the definitions of TD in ASD (RQ1.2). Section 4.3 presents an analysis of the research areas that are the focus of the literature of TD in ASD (RQ1.3), and in Section 4.4 we present the causes and consequences found in literature of incurring TD in ASD (RQ2). In Section 4.5, TDM strategies in ASD are presented (RQ3). Finally, Section 4.6 presents the research gaps identified in the area (RQ4).

### 4.1. Overview of the research of TD in ASD

There were 38 primary studies published between 2002 and 2014 that were included in the SLR. As shown in Fig. 2, there were only three studies published prior to 2010, whereas almost 97% of the studies were published in 2009 and later.

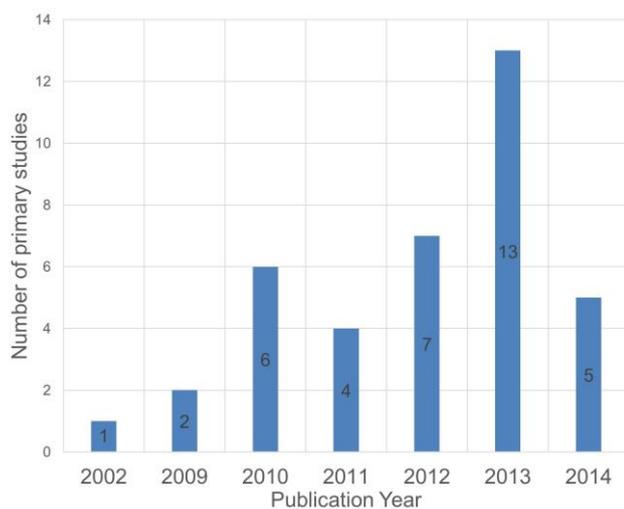

**Fig. 2. Annual distribution of primary studies**

Furthermore, when considering the distribution from 2011 to 2013, we can see that the number of publications in the area has been increasing. This trend suggests a relatively growing research interest in the topic of TD in the context of ASD. This is well-aligned with previous studies citing the growing popularity of TD in ASD (Letouzey, 2012 [P36]; Nord et al., 2012 [30]; Kruchten et al., 2013 [3]; Codabux et al., 2013 [P22]). In addition, the SLR only includes five studies published during 2014. However, the decrease in the number of studies during this period can be explained as a result of the study time period, which ended in June 2014(thereby limiting the included studies to those published in the first half of the year). The results also showed that there were no publications prior to 2002 included in the SLR, even if the inclusion criteria were set to between 1992 and 2014. One possible reason for this may be the fact that the Agile Manifesto was formulated in 2001 and that different agile methods have been acknowledged following the Manifesto.

When considering the distribution of the primary studies based on the type of venue, we found that a relatively high number of studies (84%, or 32 studies) were published in conference proceedings, while 16% of the studies (six studies) were published in journals.

We also classified the primary studies as empirical, theoretical, or both and investigated the research methods used in the primary studies. The results of the general type of research show that 69% of the primary studies (26 studies) were categorized as empirical studies, and 26% (10 studies) were theoretical papers, while 5% (two studies) were classified under both.

Regarding the research methods applied in the empirical studies, case studies were the most popular, followed by surveys. We found that there were ten case studies, seven industrial reports, five surveys, one action research study, and one experiment, as well as two studies where the research method was not stated by the authors of the primary studies.

*RQ1.1. What is the current state of the research pertaining to TD in the context of ASD in terms of research types, research contributions, and quality of the reported research?*

In order to analyze the type of publication, we used the research type classification by Wieringa et al. (2006) and categorized the primary studies as evaluation papers, solution proposals, experience reports, opinion and philosophical papers (see Fig. 3. a.). We found that 11 studies were experience papers reporting lessons learned, 10 studies were evaluation papers investigating TD in practice and conducting evaluations to validate their claims, and 9 studies were solution proposals that suggest solutions to specific problems of TD in ASD but without a full validation of the proposed solution. Opinion papers accounted for 7 studies and 1 study was philosophical paper.

Overall, the results looked promising, considering the fair distribution of evaluation and solution proposals. However, we argue that there should still be a greater focus on evaluation papers in order to validate the proposed solutions. TDM strategies that are validated in practice are needed in order to support the identification and management of TD. Based on the contribution classification by Shaw (2003) [31], research contributions were identified as shown in Fig. 3.b. Lessons learned (outcomes directly analyzed from research results) stand out first, followed by advice/implications. There were also tools (technologies, programs, or applications applied to manage TD in ASD), models, frameworks, and theoretical contributions. The findings suggest the need for more tools, frameworks, models, and other concrete guidelines that assist in understanding and managing TD in practice in ASD.

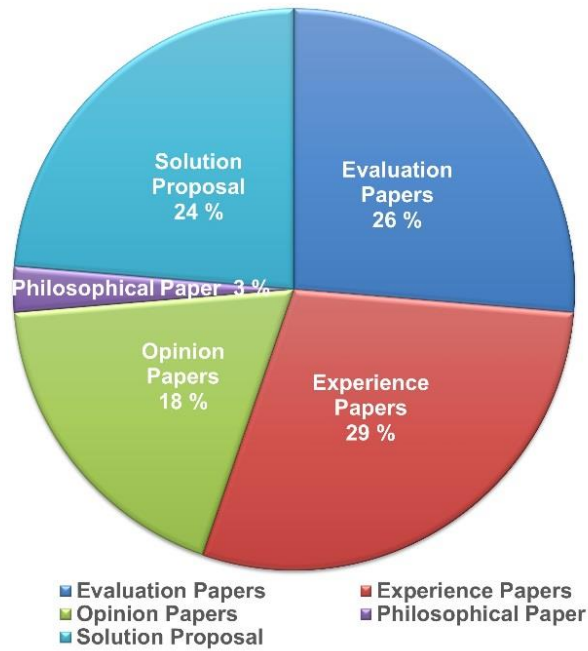
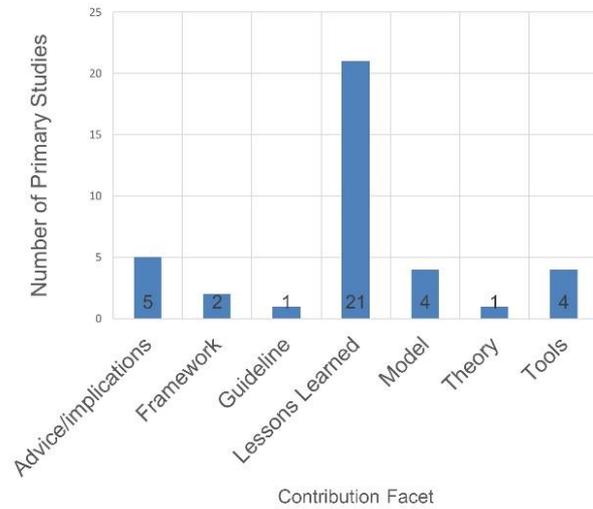

**Fig 3- (a) Distribution of research type classification**

**Fig 3- (b) Distribution of primary studies by contribution**

Finally, the results of the quality assessment of the primary studies are shown in Fig. 4. The bubble size in the figure represents the number of studies with the corresponding pair of research rigor and industrial relevance values.

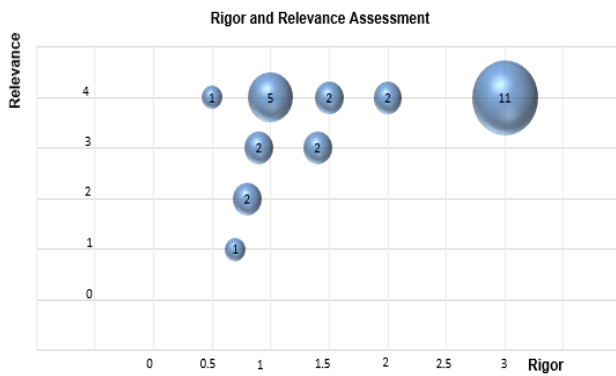

**Fig. 4. Map of research rigor and industrial relevance quality assessment**

The rigor and relevance assessment was applied to studies that drew their findings on empirical evidence. We decided to exclude the theoretical studies from this evaluation, as the Ivarsson & Gorschek (2011) [27] quality assessment model is targeted for evaluating the quality of empirical studies. Hence, the bubble plot represents only the quality assessment results from the 28 empirical primary studies included in the SLR. As can be observed from the figure, we classify 11 studies (39% of the 28 empirical studies) as high research rigor (3) and high industrial relevance (4). As a whole, there were 17 studies whose research rigor value was ≥ 1.5 and industrial relevance value was ≥ 2, and they are in the upper right corner. At the upper left position of Fig. 4, we can see that there were five studies with a research rigor value of 1 and higher industrial relevance value of 4. In addition, there were two studies that had a lower research rigor value of 1 and a favorable industrial relevance value of 3. In the lower left corner, there was one study with the research rigor value of 1 and an industrial relevance value of 1.

In general, most of our studies are highly relevant. However, 11 studies have a research rigor value equal or less than 1, which indicates the need for improving the scientific rigor of the studies conducted in the research area. These primary studies mainly lack a report on the study design and validity of the studies. According to Ivarsson and Gorschek (2011) [27], when studies with a high potential for impacting the industry (as it is the case here) fail to provide sufficient reports, they risk the chance of reduced value. Poor descriptions of the study design and validity make evaluations difficult for researchers in academia [27].

### 4.2. Aspects reflected within definitions of TD in ASD

*RQ1.2. How is the concept of TD defined in the context of ASD?*

As shown in Table 6, five aspects are emphasized in TD definitions in the context of ASD, out of which the most often mentioned is the consequences that are incurred from poor software development practices. Consequences, such as increased future cost, degraded architecture and code, reduced productivity, rigidity that impacts development and future changes, and failure to meet business demands in the long run, result from

10poor software development practices (e.g., shortsighted architectural decisions, rigidities in design and code, and shortcuts taken by developers) and are mostly considered to refer to TD in the context of ASD. This aspect of the definitions of TD in ASD is in accordance with the original definition of TD by Cunningham [1].

**Table 6. Aspects reflected in definitions of TD in ASD**

| Aspects of definitions | Primary studies | Frequency |
|---|---|---|
| Consequences that are incurred from poor software development | P1, P2, P5, P10, P17, P22, P24, P27, P30, P31, P33, and P38 | 12 |
| Deviations from design and architectural principles and coding standards | P4, P9, P11, P15, P18, P36, P37 | 7 |
| Deferred technical problems | P4, P5, P12, P13, and P25 | 5 |
| Trade-offs between short-term and long-term value | P21, P23, P24, P29 | 4 |
| Missing knowledge or inadequate (not up-to-date) documentation | P3, P14, P22 | 3 |

Deviations from standard practices of software development (that is, poor architectural, design, and code development practices) is the second most frequent aspect used to refer to TD in ASD. Unlike the first category, these definitions do not underline the associated outcomes, but rather focus on referring to TD as poor software development practices. TD has also been used to denote deferred technical problems, including technical problems or reworks that have been deferred in software development in five of the primary studies. Additionally, TD is simply used in some of our primary studies to refer to the trade-off that exists between short-term and long-term value in ASD. Missing knowledge or inadequate documentation about unspecified or underspecified requirements [P3], and inadequate documentation from incomplete user stories [P1] or documentation that is not up-to-date [P22] is also used to refer to TD in ASD.

In general, the results reveal a certain inconsistency in the use of the TD metaphor. We notice that the authors of the primary studies in our SLR use the term differently. However, we also observe that most of the TD definitions reflect dimensions of TD that have been identified in previous secondary studies on TD, such as those by Tom et al. (2013) [5], and Li et al. (2015) [6]. For example, code debt (TD that is the result of badly written code), architectural debt (TD that is caused by overlooked architectural decisions), design debt (TD caused by shortcuts taken in design), and knowledge and documentation debt (TD that occurs from the absence of well-written documentation or lack of knowledge distribution) are seen in aspects reflected in the definitions of TD in ASD.

### 4.3. Research areas of interest in the context of TD in ASD

*RQ1.3. What research areas are emphasized in the literature reporting studies of TD in the context of ASD?*

We used the keywording process described in Section 3.5 in order to identify the research areas emphasized in the literature of TD in ASD. We found that five main categories or research areas are discussed in the primary studies, as shown in Table 7. Managing TD in ASD is the most popular research area, followed by architecture in ASD and its relationship to TD. In the following, we further describe the research areas.

*4.3.1 Managing TD in ASD*

Research focused on managing TD has been prominent in ASD to meet the desire to deliver value quickly, reduce rework costs, and enhance architecture flexibility. In our study, 28 of the primary studies made reference to managing TD in ASD. Generally, this category includes primary studies that focus on the identification, measurement, monitoring, and reduction of TD in ASD. In the case of ASD, TDM requires understanding the factors that are associated with TD, such as the emphasis on quick delivery, sub-optimal architectural decisions, and solutions that balance expediency, customer value, stability, and flexibility in development. Thus, research into solutions that help conduct formal analysis and decision-making related to TD ([P1], [P5], [P8]) and into tools and models that support visualizing TD ([P9], [P20], [P25], [P37]) are emphasized in the TDM research in ASD.

**Table 7. Research areas**

| Category | Description | Primary studies | Frequency |
|---|---|---|---|
| 1. Managing TD in ASD | Identification, measurement, tracking, and reduction of TD in ASD. Practices, tools, and models used or proposed to manage TD in ASD and experience papers of TDM in ASD. | P1, P3, P4, P5, P6, P7, P8, P9, P10, P12, P14, P15, P16, P17, P20, P21, P22, P23, P24, P26, P27, P28, P29, P30, P32, P36, P37, P38 | 28 |
| 2. Architecture in ASD and its relationship to TD | The role of architecture in ASD and its relationship to TD. | P5, P7, P8, P17, P18, P21, P25, P31, P35 | 9 |
| 3. TD know-how in ASD | Understanding and assessing the know-how of TD in ASD. | P11, P22, P33, P38 | 4 |
| 4. TD in rapid fielding (expedited development) | Factors for enabling rapid fielding, where managing TD is one of the main concerns. | P2, P13 | 2 |
| 5. TD in distributed ASD | TD in the context of distributed ASD. | P11, P19 | 2 |

Models that quantify TD, such as the constructive cost model (COCOMO) [P5] and the propagation cost model [P8], as well as practices and methods to identify, monitor, prevent, and reduce TD ([P27], [P28], [P30], [P32], and [P38]) have also been focuses in the area of TDM in ASD. In this category, most of the studies were concerned with providing TD reduction strategies ([P5], [P12], [P14], [P15], [P16], [P27], [P28], [P30], and [P37]).

When considering the research type, industrial experiences that reported lessons learned in TDM of ASD projects ([P4], [P6], [P10], [P12], [P14], [P15], [P16], [P28], [P30], and [P38]) are popular. These studies were conducted to understand and manage TD in different agile team settings. For instance, P30 reports the experience of an XP team in the reduction of TD. By introducing the code Christmas tree (this method employs a visual diagram where colored squares in a chart are used to show the level of unit test coverage areas of different classes, and a second chart in the tree uses the cyclomatic complexity of the code to identify how simple and understandable the code under development is) and encouraging communication, the team was able to improve awareness of the concept of TD and its level of manifestation within the team.

Additionally, this category includes solution proposals where TDM tools and models are the focus. In P9, the DebtFlag tool is introduced for capturing, tracking, and resolving TD in software development. The DebtFlag tool is mainly comprised of an Eclipse plugin that is used for capturing, monitoring, and managing TD during development, as well as a linked web application that provides a dynamic list of TD. The authors claim that their approach follows a mechanism that supports software development methods such as Scrum. The tool maintains a TD log by capturing observations (e.g., developers identify deviations from the active requirements and create a TD list in the TD log), which can be used, for instance, in Scrum's sprint planning to define new backlog items. The tool utilizes human-made observations to capture TD during the implementation level, which enables developers to become aware of the level of TD and take action.

Considering the fact that most of the studies in this research area were lessons learned, the amount of guidelines, frameworks, tools, and models that support TDM in ASD is insufficient. A more detailed description and classification of studies in this group is presented in Appendix C. The concrete strategies to manage TD in ASD are analyzed in Section 4.5.

*4.3.2 Architecture in ASD and its relationship to TD*

Architecture in ASD and its relationship with TD is the second most popular research area in our primary studies, and nine primary studies focus on this area. Agile methods are adaptive to changes and avoid traditional software architecture that emphasizes big upfront planning, which may lead to increased documentation [P7]. Instead, agile methods rely on incremental architecture that evolves through time [P18], which may increase the risk of a lack of soundness. In ASD, software architecture inflexibility and untimely architectural decisions incur TD. Primary studies in this research area investigated the relations among software architecture, ASD, and TD. The emphasis is on combining architecture flexibility and agility in ASD ([P7], [P8], [P18], [P31] and [P35]). Understanding the role of architecture in ASD and its relationship with TD, and making informed architecture decisions that are aligned with ASD values (e.g., responding to changes), are considered essential. For instance, Brown et al. [P18] show that in ASD, the over and under anticipation of architecture is related to delayed reworks and increased costs (TD). The authors raise the need for "just enough" in ASD, focused on a balanced, informed, and flexible architecture. They suggest an agile release planning that combines the real options concept and TD to enable architecture agility as well balance flexibility. The real options concept, by comparing the business value of immediate and delayed architectural decisions, is considered to make flexible architectural decisions that balance TD in ASD [P18].

Another study, [P7], shows the importance of timely architectural decisions in ASD. As the software architecture of agile projects evolves through time, making changes in later phases can become complex and expensive. According to the authors, applying architectural decisions too early minimizes flexibility in development, especially in agile development teams.

Some of the studies utilize the relationship between architecture flexibility and ASD to analyze and support architectural decision-making related to TD ([P5], [P8], and [P17]). Fernandez-Sánchez et al. (2014) [P17] provide a model (MAKEFLEXI, Making dEcisions about Flexibility investment in Software Architecture) to assist architectural flexibility decisions in ASD. The model uses TD to bring visibility to additional costs arising from lack of architectural flexibility in ASD. Using the model, software architects in ASD can make flexibility decisions based on economic considerations (assessing the right time to either delay or invest).

*4.3.3 TD know-how in ASD teams*

The third most studied research area focuses on analyzing the know-how of TD in ASD teams. Generally, these studies are concerned with assessing practitioners' knowledge of TD, and understanding which agile practices and processes practitioners find vulnerable for incurring TD in ASD. The studies also investigate how agile teams characterize TD and perceive the consequences of incurring TD. For instance, Holvitie et al. (2014) [P33] investigate individual perceptions of TD and its manifestations among agile practitioners. In their study, the authors asked for assumed perceptions of TD among agile developers and later provided the respondents with TD definitions and its related effects, to compare the level of perception of TD with the provided definitions. Their findings indicate a difference in how practitioners perceive TD. While many respondents had prior knowledge of TD, over 20% had poor or no knowledge of the concept. The authors in [P33] argue that the concept is underutilized. Regarding the manifestations of TD, inadequate architecture was reported as the main cause for TD in their projects. The findings from the primary studies in this research area suggest that a better understanding of TD in ASD teams would help determine and clarify the scope, characteristics, and other implications of TD in ASD. An experience report of agile teams [P38] shows that the lack of knowledge of TD among agile developers contributed to reworks, poor architecture and quality, and delays that affected the relationships with customers. The authors show that raising awareness of TD among members of agile teams plays a significant role in managing TD in ASD.

*4.3.4. TD in rapid fielding (expedited system development)*

We found two studies that aim to understand factors associated with rapid fielding where TD was identified as a concern. Rapid fielding provides immediate responses to organizations interested in faster time to market and product delivery. The emphasis on quick delivery often leads to solutions that incur TD in the long run.

In rapid fielding, TD is considered as a significant factor in determining the balance between speed and stability while delivering business value. Thus, rapid fielding research with the aim of identifying solutions that balance rapid delivery and quality development and minimized TD were the main research interests. For instance, in [P2], monitoring TD with a focus on quality is revealed as an enabling factor for expediting systems development in agile projects. Development teams identified the lack of TD measures in order to convince business stakeholders to make investment in paying down TD. The study shows that a lack of TD measures made it hard for development teams to make a strong case for the business side to invest in TD fixes. Additionally, practices followed by agile practitioners to respond to problems associated with rapid fielding, such as combining architecture and agile practices, are also presented. Koolmanojwong and Lane (2013) [P13], which aims to determine factors for rapid fielding systems development, shows unmonitored TD as an inhibiting factor in rapid fielding.

*4.3.5. TD in distributed ASD*

There are two studies ([P11], [P19]) where TD is raised as a concern in distributed ASD. In distributed ASD, geographical and time distances pose a challenge for managing TD. For instance, there are difficulties in communicating TD among different teams. As a result, distributed agile teams need to have a common strategy for the accumulation and management of TD.

Bavani (2012) [P11] investigates the understanding of TD and its management in distributed ASD environments, with a special emphasis on distributed agile teams and agile testing teams. The author suggests that distributed agile teams need to be aware and aligned (that is, to achieve a common understanding on how to organize and manage TD) in order to manage TD. When distributed agile teams are aware and aligned, they are able to identify TD items easily and make optimal and informed TDM decisions. Moreover, the author argues that distributed ASD and agile testing teams should track and determine the value of paying off TD. Agile testing teams should be involved in addressing not only the TD issues of the application under development, but also the TD related to the test automation design and scripts. It has to be noted that this type of TD is different from testing debt (e.g. lack of software tests, deficient unit tests, etc.) but refers to TD in test automation scripts.

In the second study [P19], the experience of distributed ASD development on a cloud-based platform revealed TD as a key risk resulting from miscommunication among distributed teams. Once they began making changes to the code base of the cloud-based platform, distributed ASD developers at one site were unable to see the impacts on other linked parts of the code base.

Regarding the value from distributed ASD, such as production cost savings and quick time delivery [33], examining TD in distributed ASD becomes important. Understanding and internalizing the concept of TD in distributed ASD will provide the knowledge to develop approaches that enable distributed ASD that is economical and of acceptable quality. Hence, more studies should be conducted to investigate the effects of TD in distributed ASD.

*4.4. Causes and consequences of TD in ASD*

RQ2. What are the related causes and consequences of accruing TD in ASD?

We identified eight categories of causes behind incurring TD in ASD (shown in Table 8), from which the emphasis on quick delivery in ASD was identified as one of the most frequently mentioned causes. Agile developers frequently face the pressure of dealing with tight schedules while trying to deliver value to customers. As a result, they are often forced to take shortcuts in the pursuit of quick delivery, which leads to incurring TD.

Likewise, architectural and design issues such as inflexibility in architecture, poor design, and suboptimal up-front architecture and design solutions are reported by 16 primary studies as the other significant cause behind incurring TD in ASD. While this can probably be attributed to the shortsighted view of architecture within ASD, it also indicates the need for balancing architecture and agility in ASD. We need approaches that deliver robust and flexible architecture that enables flexibility and adapting to dynamic business changes.

When ASD practices such as test automation are not properly followed, they can result in TD. There are 10 studies where inadequate test coverage is reported as a cause of incurring TD. The lack of automated tests, acceptance tests, and integration tests contributes to testing-related TD. Similarly, nine studies attributed TD in ASD to the lack of understanding of the systems being built (requirements). A failure to understand the system under development may lead developers to make assumptions that consequently incur TD in ASD.

Table 8. Causes of incurring TD in ASD

| Cause | Primary studies | Frequency |
|---|---|---|
| Emphasis on quick delivery | P1, P2, P4, P5, P8, P11, P19, P21, P23, P24, P26, P27, P28, P34, P35, P38, | 16 |
| Architecture and design issues | P2, P4, P5, P10, P12, P13, P17, P18, P21, P25, P27, P31, P33, P35, P36, P38 | 16 |
| Inadequate test coverage | P5, P6, P11, P12, P16, P23, P27, P28, P32, P38 | 10 |
| Lack of understanding of system being built/requirements | P5, P7, P11, P13, P19, P20, P27, P28, P29 | 9 |
| Overlooked and delayed solutions and estimates | P5, P7, P10, P11, P13, P21, P27, and P28 | 8 |
| Less/no/delayed refactoring | P5, P8, P21, P23, P28, P37 | 6 |
| Code duplicates/copy pasting | P11, P16, P29 and P38 | 4 |
| Others | P5, P22 | 2 |

Another significant factor associated with TD in ASD is an oversight in estimations of sprints in ASD and subsequent delayed decisions. Overlooked estimations in sprints, schedules and working velocity, delayed refactoring, and architectural decisions incur TD in ASD. Unrealistic estimations and delayed decisions can be the result of inexperienced ASD teams. However, the literature suggests that these can be mitigated by improving team's estimation ability [P10].

Inadequate refactoring or delayed refactoring is also reported by six studies as a cause of incurring TD in ASD. This implies that developers should give enough attention to refactoring in order to prevent TD from building up in the software product. Another significant cause of incurring TD came from code duplicates (copy pasting), which were implemented as shortcuts during development.

Furthermore, we identified other non-recurring causes related to incurring TD in ASD, such as: parallel development in isolation where the buildup of TD increase when source code from isolated branches are merged to the code base [P5], resource constraints [P22], and an organizational gap between business, operational, and technical stakeholders [P22].

While considering the consequences of incurring TD in ASD, reduced productivity, system quality degradation, and increased maintenance costs are the most notable ones (see Table 9).

Table 9. Consequences of incurring TD in ASD

| Consequence | Primary studies | Frequency |
|---|---|---|
| Reduced productivity | P1, P2, P5, P8, P10, P11, P13, P14, P16, P17, P20, P23, P29, P31, P32, P37, P38 | 17 |
| System quality degradation | P3, P4, P5, P7, P8, P10, P13, P15, P20, P22, P25, P26, P27, P32, P33, P37, P38 | 17 |
| Increased cost of maintenance | P8, P11, P12, P13, P17, P18, P21, P22, P23, P24, P26, P27, P29, P30, P38 | 15 |
| Complete redesign or rework of system | P13, P35, and P38 | 3 |
| Market loss/ hurt business relationships | P11, P20, P38 | 3 |

Whenever TD is not incurred strategically, it builds up to a level that forces agile teams to put a great deal of effort into fixing defects and addressing stability issues. This in turn results in a slowdown of working speed and reduced productivity. Similarly, TD affects the quality of a system through degradation, making it error-prone, less stable and at times redundant ([P5], [P22] and [P13]).

The other significant consequence of TD in ASD was the increased cost of maintenance. Failure to address TD on time accrues more TD and results in a more complex and hard to maintain software with an increased cost of maintenance ([P8], [P11], [P12], [P13], [P17], [P18], [P21], [P22], [P23], [P24], [P26], and [P27]). In such cases, agile teams need to put in extra effort to fix sub-optimal design choices, and the further introduction of post-delivery costs significantly affects the revenue that could be gained if there were proper management of TD.

Complete redesign or rework of the system as well as a market loss have also been reported as consequences of TD in ASD. From the developers' perspective, TD forces big bang responses, such as taking shortcuts in maintenance, which further increase the accumulation of TD ([P2] and [P4]). A loss in traceability or predictability might also arise, forcing either a complete redesign or rework of the system ([P13], [P35], and [P38]). When left unmanaged, TD can grow to a level that is more complex and costly. TD in ASD can also lead to a market loss and failure of businesses ([P11], [P20]) or negatively impact the relationship with business stakeholders [P38]. In contrast, when TD is incurred systematically, it can be used to gain business value and opportunities and to achieve project goals ([P5], [P10], and [P21]).

*4.5. TDM strategies in ASD*

*RQ3. What are the strategies proposed in literature to manage TD in ASD?*

We identified 31 studies addressing RQ3, from which 12 categories of TDM strategies in ASD were derived. As shown in Table 10, there are 12 studies where the use of specific approaches, tools, and models are suggested for managing TD in ASD. In general, these are TDM strategies that assist with decisions in ASD, tools for identifying, tracking, and resolving TD in ASD, and models that help in the analysis of TDM decisions.

For instance, Cavaleri et al. (2012) suggest in [P23] that having a project problem-solving pattern in ASD would help to manage TD concerns, as well as measure and monitor TD. Project problem-solving patterns are "*the recurring organizational configuration of human interactions designed to support a project team in seeking, recognizing and formulating problems*" [P23]. Others such as Holvitie et al. (2013) [P9] proposed the DebtFlag tool for identifying, tracking, and managing TD in ASD.

Our findings reveal that refactoring (a TD repayment strategy) is the most popular TDM strategy in ASD. Refactoring helps to pay down TD by restructuring the code base or system architecture without altering the external behavior of the software system under development. Suggestions regarding the refactoring strategy for TDM in ASD include educating developers to refactor, embracing refactoring culture, and applying continuous refactoring ([P4], [P13], [P18], [P20], [P21], [P22], [P24], [P27], [P28,] [P32], and [P33]). However, there are some concerns with refactoring in ASD. For instance, Nord et al. (2012) [P21] argue that refactoring may not be as effective in large-scale projects as it is in small-scale projects. Additionally, it is considered time-consuming and hard to apply in complex projects [P32]. TD quantification is also key in managing TD in ASD. We found cost models such as the COCOMO ([P5], [P17]), project solving pattern framework [P23] and the propagation cost model [P8] were used to quantify TD and analyze decisions related to TD in ASD. The detailed list of approaches, tools, and models used for managing TD is shown in Table 11.

**Table 10. TDM strategies in ASD**

| TDM strategy | Primary studies | Frequency |
|---|---|---|
| 1. Specific approaches, tools and models to manage TD in ASD | | 12 |
| • Tools | P9, P11, P25, P32, P38 | 5 |
| • Approaches | P1, P7, P21, P23 | 4 |
| • Models | P5, P8, P17 | 3 |
| 2. Refactoring | P4, P13, P18, P20, P21, P22, P24, P27, P28, P32, P33 | 11 |
| 3. Enhanced visibility of TD | P2, P7, P8, P21, P30, P32, P36, P37 | 8 |
| 4. Test automation | P11, P16, P22, P32, P34 | 5 |
| 5. Common (agreed) DoD | P12, P14, P34, P36, P37 | 5 |
| 6. Planning in advance for TD | P4, P11, P16, P22, P27 | 5 |
| 7. Code analysis | P4, P12, P15, P16, P27 | 5 |
| 8. Agile practices such as pair programming, TDD (test driven development) and CI (continuous integration) | P6, P16, P22, P28, P29 | 5 |
| 9. Prioritizing TD | P11, P12, P27, P389 | 4 |
| 10. Improving estimation techniques | P10, P27, P28, P29 | 4 |
| 11. Transparent communication as to the level of TD with business stakeholders | P2, P23, P28 | 3 |
| 12. Establishing an acceptable level of TD | P4, P32, P36 | 3 |

**Table 11. Approaches, tools, and models for managing TD in ASD**

| Primary study | Approaches, tools, and models |
|---|---|
| P1 | (Approach) Cost-benefit analysis approach |
| P5 | (Model) Cost-model (COCOMO) to analyze trade space and show the range of options and the resulting consequences related to TD |
| P7 | (Approach) Responsibility driven architecture approach applying concepts of real options theory to track decision-making and understand timing for appropriate TD decisions |
| P8 | (Model) Propagation cost model to get an insight into degrading architecture quality |
| P9 | (Tool) DebtFlag tool to capture, track, and resolve TD |
| P11 | (Tool) Continuous integration tool to quantify TD on the basis of code complexity and automated test coverage |
| P17 | (Model) MAKEFLEXI and COCOMO models to estimate and manage TD in ASD |
| P21 | (Approach) Collective dashboards and visualization approaches (e.g., assisting TDM through a sonar visualization plugin) |
| P23 | (Approach) Project problem-solving pattern approach |
| P25 | (Tool) Dependency analysis and revision history to detect architectural deviations |
| P32 | (Tool) Quality "from now" through tools such as Ndpend to detect and monitor code violations |
| P38 | (Tool) TD board to visualize and manage TD, accompanied with metrics to measure TD (static code analysis tools) |

Enhancing the visibility of TD, architectural dependencies, and a list of design decisions related to TD are reported as TDM strategies in ASD by eight studies ([P2], [P7], [P8], [P21], [P30], [P32], [P36], and [P37]). Improving TD visibility by keeping track of a list of architectural and design decisions in a backlog ([P2], [P7]), the use of TD visualization boards ([P21], [P36], and [P38]), " code Christmas tree" [P30], pie and bar charts to visualize and manage TD [P37], and TD visualization tools like Ndpend to detect code violations [P32] that assist in identifying, tracking, and managing TD, are included in this category. Increased visibility of architectural dependencies is argued to enhance rework estimations and communication of architectural quality ([P8], [P21]).

Test automation has also been reported as a strategy to reduce the level of TD incurred in ASD ([P11], [P16], [P22], [P32], and [P34]). This is done by automating manual tests or assigning test automation teams [P22] to increase the coverage of automated tests. However, it is important that agile teams understand and apply test automation practices properly. Agile developers and testers should collaborate in the design of automated test scripts [P11].

The "Definition of Done" (DoD) concept was also commonly applied as a TDM strategy in ASD. Establishing a common understanding on what "done" means (e.g., Scrum's "Definition of Done") within agile teams has a significant value in TDM. The concept of DoD has been applied to improve the quality of software and reduce TD resulting from deferred defects in five studies ([P12], [P14], [P34], [P36], and [P37]). Employing a common DoD in different levels, such as story, sprint, and release, in order to achieve a common understanding on TD-related issues and strategically manage TD is a unique TD reduction strategy applied in ASD. For instance, in P12, multi-level DoD (DoD at the story, sprint, and release levels) is used to monitor TD with metrics such as deferred defects (measure of TD for future releases) and reopened defects. A DoD with fewer deferred defects shows reduced TD. The strategies related to "Common DoD" reported in the primary studies are shown in Table 12.

**Table 12. Common DoD for managing TD in ASD**

| Primary study | Common (agreed) DoD |
|---|---|
| P12 | Multilevel DoD (DoD at story, sprint, and release levels) is used. |
| P14 | DoD is used during sprint reviews to identify and highlight fall out items (incomplete stories). |
| P34 | DoD is applied (through quality assurance checklists) in order to prioritize TD. |
| P36 | "Definition of the right code" is applied to manage TD. |
| P37 | TDM is included as part of the DoD for user stories, sprints, and releases. |

Planning ahead for TD is indicated as a TDM strategy in ASD in five studies ([P4], [P11], [P16], [P22], and [P27]). Advance planning for TD involves assigning pre-emptive efforts to address TD. Resources were allocated in advance for tasks such as code cleanup and the removal of design shortcuts [P4], code review and refactoring [P27], and more appropriate planning of user stories to pay down TD in the future [P11]. Dedicated teams that are particularly responsible for TD reduction can also be allocated in advance in agile teams ([P4], [P16], and [P22]).

Additionally, code analysis through code review, test reviews, and the use of automated tools were also reported as important practices for managing TD in ASD, as shown in Table 13. These cover manual and automated code reviews and code analysis practices help to identify deviations from coding conformance and testing.

The agile practices of pair programming, test-driven development, (TDD), and continuous integration (CI) are also reported as strategies for managing TD in ASD in five studies ([P6], [P15], [P21], [P27], and [P28]). For instance, with CI, developers were able to check whether the integration of new code broke the existing code base and make corrections quickly ([P6], [P28]). Applying TDD is also reported as a practice for managing TD in small-scale agile software projects ([P15], [P21], and [P27]). Furthermore, through pair programming, developers can easily communicate and learn about TD and its management [P27].

**Table 13. Code analysis for managing TD in ASD**

| Primary study | Code analysis |
|---|---|
| P4 | Code reviews to reveal design shortcuts. |
| P12 | An acceptance test peer review and design peer review. |
| P15 | Code review with better-defined acceptance criteria. |
| P16 | Application of automated code analysis (code coverage, code, and design rule conformance). |
| P27 | A review to understand where the code went wrong. |

We also identified four studies ([P11], [P22], [P27], and [P38]) that suggest prioritizing TD as a TDM strategy in ASD projects. TD in ASD needs to be classified and ranked according to factors such as severity during the process of paying it down. For instance, Davis (2013) [P12] proposes prioritizing TD items in earlier iterations of ASD/flows of lean development as a preventive measure to pay down TD in ASD.

There are four studies where improving estimation techniques (e.g., tackling estimation problems) is found as a mitigation strategy for TD in ASD projects (see Table 14). Optimistic estimations by agile developers lead to TD ([P10], [P28]). In such cases, increasing the teams' estimation ability leads to improved size estimates and incurs a smaller amount of TD.

**Table 14. Improving estimation techniques in ASD**

| Primary study | Improving estimation techniques in ASD |
|---|---|
| P10 | Improve the agile team's estimation ability. |
| P27 | Empower teams in estimation; agile teams should keep realistic estimates (hours for code review and refactoring). |
| P28 | Practice frequent deadlines in previous user stories, which is assumed to improve estimates for new user stories. |
| P29 | Develop a method for bulk estimating a release backlog to identify high-risk backlog items. |

While ASD promotes engaging customers in the product development, establishing a transparent communication about TD with customers can be a concern. Developers find it difficult to communicate a business case for TDM decisions with business stakeholders [P2]. In order to fulfill the continuous demand of customers, agile developers might focus on speedy delivery and make suboptimal decisions that incur TD. Moreover, failing to communicate TD transparently and pay it back at the right time impacts the team's productivity and schedules [P2]. There are three studies where building transparent communication as to the level of TD with business stakeholders is observed as a strategy towards managing TD in ASD ([P2], [P23], and [P28]). By communicating the level of TD clearly with customers, agile developers can collaborate with customers to make appropriate decisions that balance delivery as well help manage TD.

We identified three papers which discern establishing a consensus about the acceptable level of TD as a TDM strategy in agile projects ([P4], [P32], and [P36]). These studies show that it is important for teams to establish an agreement on the minimum quality of the code base and the level of TD that can be assumed. It was observed that by agreeing on the acceptable level of TD, agile teams can prevent further increase of TD and take corrective action in time [P32]. A consensus as to the minimum quality of source codes added to the code base helps to keep TD to a minimum [P4].

### 4.6. Research gaps

*RQ4: What are the existing research gaps in the field of TD in ASD?*

The findings of our study provide important research areas that need further investigation. We use the relationship between facets such as pertinence, research type, and contribution to analyze and explore research gaps in the area. As the first step, we wanted to investigate the relationship between the pertinence and contribution facets of primary studies, as shown in Fig. 5.a. Information for determining the pertinence, contribution and research type of studies is shown in Appendix B.

By analyzing the pertinence and contribution facets of the studies in Fig. 5.a, it can be observed that only 10 studies have full pertinence (entirely dedicated to discussing TD in ASD). However, among these studies, eight have a lessons learned contribution; the rest include one tool and one guideline contribution. This further reflects a scarce contribution of models, frameworks, theories, guidelines, and tools that are fully focused on TD in the context of ASD. Hence, we can argue the need for more studies that are fully pertinent and offer such a necessary contribution to help characterize and manage TD in ASD. Additionally, Fig. 5.a reveals that 17 studies have partial pertinence, and the other 11 studies have marginal pertinence. We also undertook an investigation of the research type to determine the trends through the years, as shown in Fig. 5.b. Philosophical studies constitute the least employed research type in the area. Given that philosophical studies introduce a new and innovative way of looking at things, we can also argue the need for more philosophical papers that provide novel approaches to characterize and manage TD in ASD.

In addition, while considering publications from 2011 onward, we can see that there has been more emphasis on evaluation papers (10 studies). It can also be seen that there are six experience papers, eight solution proposals, four opinion papers, and one philosophical paper during the same period of time. A relatively growing adoption of evaluation studies during this period shows that there have been more practical implementations and their evaluations conducted in the area. The distribution also shows that a relatively high number of studies (about 24%) in the area are solution proposals. The increasing adoption of evaluation and solution proposal papers should further be encouraged

Additionally, we observe concrete research gaps that need further investigation, which are presented as follows.

The two most important research areas imply that academia should focus on providing solutions for managing TD in ASD and further investigate the role of architecture in ASD and its relationship with TD. For instance, there is a need for solutions that quantify degrading architecture quality (architectural debt) [P8] and characterize the business value of architectural decisions in ASD.

Considering the different tools and models for managing TD in ASD, we need standardized, advanced, and validated approaches, tools, and models. We found studies that recommend tools, models, and approaches without empirical validation. For instance, the DebtFlag tool suggested in [P9] supports only the Java language and lacks cross-compatibility (that is, support for other programming languages). Moreover, there is no empirical validation of the tool.

Immediate delivery of software should counterbalance the delivery of overall business value and likely accrued TD [34]. In the context of rapid fielding, solutions that balance schedule (expediency), cost, and flexibility are required for managing TD. We need validated cost models that support a method of TDM decision-making that values both expediency and quality in ASD.

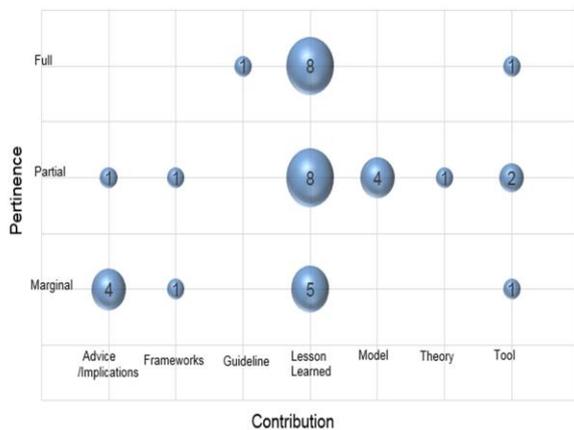

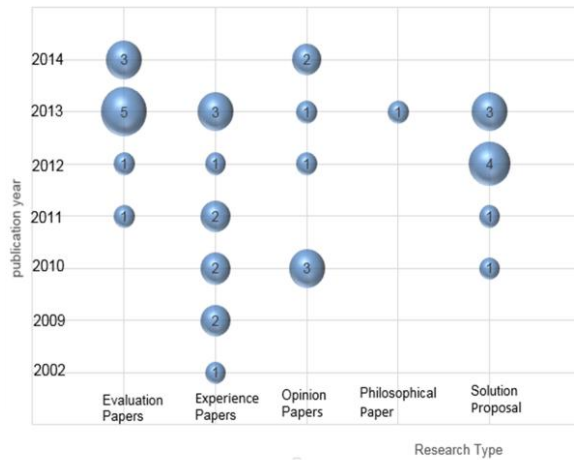

**Fig. 5 - (a) Contribution Vs pertinence of primary studies**  **Fig. 5 - (b) Annual distribution of studies by research type**

Similarly, a detailed investigation of TDM strategies, such as the applicability of refactoring in large-scale ASD, remains to be done ([P21], [P24], and [P32]). The lack of support mechanisms to aid developers in paying down TD through refactoring was reported as a challenge [P32]. Future studies can examine ways to adapt refactoring in complex and large-scale agile projects and provide refactoring decision-making approaches that support managing TD in ASD.

In ASD, strategies that clarify the business value of TDM decisions, especially from the business stakeholders' perspective, are critical for communicating the state of TD. While TD problems may be visible for developers, communicating the impact of TD with business stakeholders is a challenge [P2]. Identifying strategies that increase the business visibility of TD [P10], provide transparent communication on the value of TD, and assist in TDM decisions among developers and business stakeholders is a potential research gap that needs to be addressed. Although none of our primary studies explicitly mentioned the concept of "social debt", aspects that reflect social debt were found for example in social interactions and organizational structure that are affected by TD. For instance, there were cases where the lack of trust and informal communications were affecting decisions related to TD in the case of distributed software development [P19]. Cases related to organizational gaps that affect TD have also been discussed. [P22]. Therefore, social debt in ASD seems to be a fruitful area for future research.

Regarding the application of DoD in ASD, some empirical studies show its effectiveness in managing TD incurred from code defects. However, the implication of DoD for non-functional requirements is still an open research issue, and validations are necessary to determine a relevant DoD strategy [P12].

Finally, the majority of the research contribution in the area is lessons learned. Therefore, academia should emphasize providing more tools, models, and frameworks to assist formalization and management of TD in ASD.

## 5. Comparison to related work

Our SLR aggregates the existing knowledge on TD in the context of ASD in terms of the most studied research areas, the causes of TD in ASD, the consequences of incurring TD in ASD, and strategies for identifying, measuring, and preventing TD in ASD. We included a significant number of primary studies that have not been part of the previous secondary studies ([P5], [P7], [P8], [P14], [P15], [P17], [P18], [P19], [P20], [P23], [P25], [P26], [P28], [P31], [P32], and [P35]).

While the context of our research (ASD) is completely different from the other studies, we observe close relationships between the research questions of existing secondary studies and our SLR. Table 15 provides the list of corresponding questions raised in other studies in comparison to ours.

For instance, regarding the causes of incurring TD in ASD (RQ2), we found eight categories of causes (see Table 8). Similarly, Tom et al. (2012; 2013) ([4], [5]) investigated the reasons behind TD, asking "why does technical debt arise?" The authors highlight the precedents of TD as a prioritization of delivering the product (time constraint), poor communication, poor collaboration among team members, poor documentation, and individual attitudes such as ignorance and oversight [5]. Among those, the first cause (emphasis on quick delivery) is also considered to be one of the most common potential causes of TD in ASD. Our analysis shows that architecture and design issues are also the most common causes of TD in ASD. Our work complements the findings of Tom et al. (2012) and introduces new perspectives regarding the causes of TD in ASD, such as overlooked estimations of sprints, parallel development in isolation, issues in testing and coding processes, and organizational gaps among business, operational, and technical stakeholders. Alves et al. (2015) did not study the causes of TD, but the authors looked at indicators to find different types of TD. Under the design and documentation debt categories, Alves et al. (2015) highlight bad coding practices, architecture issues, and documentation issues as potential reasons for incurring TD, and propose a list of indicators from coding and design to identify TD (e.g., modularity violation, code smells, and automatic static analysis). In our SLR, we also found coding and design issues (e.g., code duplicates and insufficient or no refactoring) to be potential causes of TD in ASD. Architecture and design issues were also among the popular themes, but coding debt was not popularly reported by primary studies in the context of ASD.

**Table 15. Corresponding questions with other secondary studies**

| Secondary study | Research questions | |
|---|---|---|
| Our study | RQ2. What are the related causes and consequences of incurring TD in ASD? | RQ3. What are the strategies proposed in literature to manage TD in ASD? |
| Tom et al. (2012;2013) | RQ2. Why does technical debt arise? RQ3. What are the benefits and drawbacks of allowing TD to accrue? | |
| Li et al. (2015) | | RQ6. What are the different activities of TDM? RQ7. What approaches are used in each TDM activity? RQ8. What tools are used in TDM and what TDM activities are supported by these tools? |
| Alves et al. (2015) | | Q3. What strategies have been proposed for the management of TD? |
| Ampatzoglou et al. (2015) | | RQ2. What are the financial approaches that have been applied in technical debt management? |

Regarding our second research question (RQ2), we can infer that in ASD the focus should be on balancing the immediate and long-term concerns to minimize TD. More specifically, resource constraints and architecture and design issues are the causes driving TD in ASD, and have similarly been reported by Tom et al. (2012; 2013) and Alves et al. (2015) as two main factors that in turn lead to TD.

In our SLR, reduced productivity, system and quality degradation, and increased maintenance cost were found as the most significant consequences of incurring TD in ASD. Likewise, the study by Tom et al. ([4], [5]) of the outcomes of TD in software development reveals that TD impacts productivity, the quality of the product, and project risk. Moreover, Tom et al. (2012) show that team morale is impacted as a result of incurring TD. While we did not find consequences related to team morale, other consequences (such as market loss and a complete redesign/rework) were the other recurrent themes in our findings. Therefore, we could argue that the consequences of TD are common among software teams regardless of the development approaches implemented.

Regarding the TDM strategies in ASD (RQ3), we find relevant questions raised by Li et al. (2015) [6], Alves et al. (2015) [7] and Ampatzoglou et al. (2015) [8] despite the difference in context. Li et al. [6] investigate TDM in terms of identification, measurement, repayment, communication, and prevention techniques. The authors suggest both code and dependency analysis to identify TD. In our SLR, we also looked for specific approaches, tools, and models to identify TD in ASD and found that code and architectural dependency analysis are techniques suggested in the primary studies. Different from earlier work, detailed strategies are proposed in the context of ASD to identify TD, such as acceptance test reviews, code and design reviews, and automated analysis. To measure TD, Li et al. [6] identified cost estimation models, human estimation, and code and operational metrics. In our SLR, we also find that the COCOMO model, propagation cost models, project solving patterns and the more specialized MAKEFLEXI cost model are different ways to measure and monitor TD in ASD. In terms of agile practices, we observe that collective dashboards, visualization techniques, and continuous integration tools were also proposed in the primary studies to identify and monitor TD in ASD. To reduce TD, Li et al. [6] propose refactoring, automation, reengineering, and bug fixing activities; among these, refactoring, code analysis, and test automation are also reported in our SLR. Furthermore, we report more agile-specific approaches to monitor and reduce TD, such as setting a commonly agreed DoD, improving estimation techniques of sprints, planning in advance for TD, and implementing pair programming or test-driven development.

Ampatzoglou et al. [8] investigate financial approaches used in TDM. Similarly, Alves et al. (2015) summarize TDM strategies. The authors in both of these studies list cost-benefit analysis, portfolio management, and real options as frequently applied techniques. Ampatzoglou et al. [8] mention that these financial approaches are employed differently in various studies, and there is an inconsistency between their use in software engineering and finance contexts. Alves et al. [7] confirm that the same approaches were employed for TDM, but they further add both the analytical hierarchical process and dependency analysis as more software–engineering specific approaches. In our SLR, we also find cost-benefit analysis as one strategy for TDM in ASD. Furthermore, we identify a list of primary studies employing custom approaches that have not been listed in previous secondary studies, e.g., a responsibility-driven architecture approach (based on real options theory) and a project problem-solving pattern. These custom approaches also confirm Ampatzoglou et al. [8]'s findings that TDM has been done differently in different settings, and there is a lack of consistency in this regard.

Overall, we observe that studies on ASD share similar causes of TD (e.g., coding and architecture issues), encounter similar consequences of incurring TD (e.g., reduced productivity and quality), and the studies employ similar strategies for TDM (e.g., dependency analysis, automated static code analysis, testing activities, cost estimation models, and refactoring). Our study, on the other hand, complements prior work on TD by introducing TDM strategies used in agile teams (e.g. 'common DoD', improving estimation techniques for sprints), causes of incurring TD in ASD (e.g. oversight in sprint estimations, isolated parallel development), and new approaches, tools, and models for identifying, measuring, and reducing TD in the specific context of ASD (e.g. MAKEFLEXI,. project solving pattern).

## 6. Conclusion

TD has broad economic and technical implications in ASD, and has recently been gaining more attention from both academia and industry. As a result, an increasing number of studies have been conducted on TD within the context of ASD. Our study yielded 38 primary studies discussing TD in the context of ASD.

In this study, we determine five research areas from the literature of TD within the context of ASD. These are: managing TD in ASD, architecture in ASD and its relationship with TD, TD know-how in ASD teams, TD in distributed ASD, and TD in rapid fielding development. Among these research areas, great attention was given to managing TD in ASD.

The majority of literature discussing TD in the context of ASD reports consequences of poor software development practices to describe TD. An emphasis on quick delivery and architecture and design issues are the most frequent causes attributed to TD within the context of ASD. The lack of understanding of the system being built (requirements) and inadequate test coverage are the second most reported causes of incurring TD in ASD. Additionally, the study identifies causes such as overlooked solutions and estimates, less/delayed refactoring, code duplicates, parallel development in isolation, and organizational gaps among business, operational, and technical stakeholders attributed to incurring TD in ASD. Regarding the consequences of TD in ASD, reduced productivity, system (quality) degradation and increased maintenance costs are the top three consequences. There are various approaches, models, and tools employed to identify, quantify, monitor, and pay back TD in in the specific context of ASD. In terms of TDM strategies in ASD, refactoring is the most popular practice used to repay TD. Additionally, a significant number of studies suggest enhancing the visibility of TD as a TDM strategy in ASD. Applying common (agreed) DoD, test automations, and planning in advance for TD are also popular strategies of managing TD in ASD.

With regards to practitioners, the results of our study provide knowledge on TD and its management within the context of ASD. They also provide a comparison with previous studies' categorization of TD concepts, highlight techniques used in general by the software engineering community, and introduce new approaches that are specifically useful for ASD. Through this study, practitioners can identify the most common reasons for incurring TD, its consequences, and the economic implications of ASD and TDM strategies applied in ASD. For instance, when employing test automation for managing TD in ASD, it is important that agile testers ensure that their test automation design and scripts will not incur TD.

From an academic perspective, the study identifies important research areas that need further investigation. Future studies should emphasize investigating TDM in ASD, as well as the role of architecture in ASD and its relationships to TD. Our findings also indicate the need for more tools, models, and guidelines that support management of TD in ASD. Moreover, there is a potential research gap for standardized approaches to manage TD. Examining the relationship between causes and consequences of TD in ASD will assist in uncovering potential TDM strategies that address TD-related issues. These findings would help agile practitioners to understand different causes of TD in ASD, monitor and prevent TD in their teams through agile specific strategies.

### Acknowledgements

This research has been partially supported by ICT SHOK N4S program financed by the Finnish Funding Agency for Technology and Innovation (Tekes) and Digile OY.

### Appendix A. Primary studies

[P34] Wiklund, K., Eldh, S., Sundmark, D., & Lundqvist, K. (2012). Technical debt in test automation. pp. 887-892.

[P35] Palmer, K. D. (2014). The essential nature of product traceability and its relation to agile approaches., 28. pp. 44-53.

[P36] Letouzey, J. -. (2012). The SQALE method for evaluating technical debt. pp. 31-36.

[P37] Power, K. (2013). Understanding the impact of technical debt on the capacity and velocity of teams and organizations: Viewing team and organization capacity as a portfolio of real options. pp. 28-31.

[P38] Dos Santos, Paulo Sérgio Medeiros, Varella, A., Dantas, C. R., & Borges, D. B. (2013). Visualizing and managing technical debt in agile development: An experience report Springer.

**Appendix B. Extracted data properties**

| Property | Description |
|---|---|
| Pr1. Publication Year | Publication year of the primary study. |
| Pr2. Publication Source | Publication forum used to disseminate the primary study e.g. conference or journal. |
| Pr3. General type of paper and research method | Studies were classified as a) *empirical* when the study findings were based on direct empirical evidence, b) *theoretical* if the study findings primarily base on understanding of a certain field but without any empirical evidence to support the findings or suggestions made in the study, and c) *both* if they were a combination of both empirical and theoretical studies.<br><br>Additionally, empirical studies were classified according to the specific empirical research method reported by the authors of the primary studies as follows:<br><br>• Case study: the study employs case study, exploratory study where researchers analyze and answer predefined questions for a single or multiple cases.<br>• Survey (questionnaire, observation, interview)<br>• Industrial report.<br>• Action Research: apply a research idea in practice, evaluate results, modify idea (cross between experiment and case study).<br>• Experiment: empirical enquiry that investigates causal relations and processes.<br>• Not stated |
| Pr4. Research Type | Research type classification adapted from Wieringa et al. (2006).<br><br>• Evaluation paper: investigates the problem in practice or techniques that are implemented in practice and evaluation is conducted to validate the knowledge claim.<br>• Solution proposal: proposes a solution or technique and argues for the relevance without full validation. The solution can be novel or significant improvement of an already existing technique.<br>• Philosophical paper: sketches or introduces new way of looking at things, a new conceptual framework, etc.<br>• Opinion paper: describe the personal opinion of the author about some topic, practice, etc. Explains what is good and bad about something.<br>• Experience paper: reports personal experience of the author on one or more projects, it explains lessons learned by the author. |
| Pr5. Pertinence of the paper | Extent to what the study discusses TD in the context of ASD.<br><br>• Full: entirely related, the paper's main focus is related to investigating TD in the context of ASD<br>• Partial: while the main topic of the research can be different from TD, the paper discusses the concept in the context of ASD to an average level.<br>• Marginal: slightly discusses issues of TD in the context of ASD, but the main topic of the research is different from TD. |
| Pr6. Contribution | Adapted from Shaw, (2003) was used to identify the type of contribution from the primary study.<br><br>• Model: representation of an observed reality by concepts or related concepts after conceptualizing the process.<br>• Theory: construct of cause-effect relationships between determined results.<br>• Framework: conceptual maps and methods that help in analyzing and managing TD in ASD<br>• Lessons learned: a set of outcomes, directly analyzed from and obtained from research results (results found from industrial/experience reports were also considered as lessons learned).<br>• Guidelines: list of advice, synthesis of obtained research results.<br>• Tools: technologies used, programs or applications applied to manage TD in ASD.<br>• Advice/implications: recommendations driven from personal opinions of the authors |
| Pr7. Research rigor | Assess rigor of the research method of an empirical study based on (Ivarsson and Gorschek, 2011) model of rigor and relevance, as follows:<br><br>1. Evaluate the extent to which 3 aspects (Context, Study design, and Validity) are described<br>• Context (e.g. description of development mode, speed, company maturity)<br>• Study design/research method (measured variables, treatments, controls used in the study)<br>• Validity (description of threats to validity, measures to limit threats<br><br>*These aspects are scored on 3 level "weak" [0] medium [0.5] strong [1]. Total rigor is calculated as the sum of the rigor values of the 3 aspects. |

| Property | Description |
|---|---|
| Pr8. Industrial relevance | Assess industrial relevance of an empirical study based on (Ivarsson and Gorschek, 2011) model of rigor and relevance, as follows:<br>1. Evaluation of the realism of the study environment:<br>• Subjects (practitioners, students, researchers)<br>• Context (industrial setting)<br>• Scale (realistic size usefulness scalability)<br>2. Evaluation of how the research method applied in the study contributes to its industrial relevance:<br>• Research method (action research, case studies, conceptual analysis, survey, interview, experiment)<br>*All these aspects are evaluated on 2 levels as "Contribute to relevance" [1] and "Do not contribute to relevance" [0].<br>Then, the total relevance is computed as the sum of the relevance values of the 4 aspects |
| Pr9. TD definitions in ASD | Definitions of TD in the context of ASD that are reported in literature. |
| Pr10. TD cause in ASDs | Identify reasons behind incurring TD in the context of ASD |
| Pr11. TD consequences in ASD | Identify the result of incurring TD in the context of ASD |
| Pr12. TDM strategies in ASD | TD management mechanisms reported in literature e.g. practices, approaches etc. |

**Appendix C. Primary studies investigating TDM, their research type and contribution facets**

| PS | TDM Focus | Research type | Contribution |
|---|---|---|---|
| P1 | TD formalization and decision making approach to help TDM that can be applied in ASD. | Philosophical | Theoretical |
| P3 | Proposes an adaptive performance modelling supported with automated performance analysis in ASD where TD is paid off strategically. | Solution proposal | Model |
| P4 | Educates and encourages to refactor, build consensus on minimum quality of source code additions, apply code review, plan in advance for TD, and assign special teams responsible for TD. | Experience paper | Lessons learned |
| P5 | Shows how engineering cost model (COCOMO) is used in the analysis of options and consequences into balance expedited engineering, increase flexibility in architecture and minimize TD. | Solution proposal | Model |
| P6 | Shows how automated testing through continuous integration enable agile testing and slowed down testing TD | Experience paper | Lessons learned |
| P7 | Shows how a balance between architecture and agility can be used to manage TD. | Opinion paper | Advice |
| P8 | Demonstrates the propagation cost analysis model (propagation cost metric to model the impact of degrading architectural quality by quantifying degrading architectural quality and the potential for future rework costs during iterative release planning to strategically manage TD). | Evaluation paper | Model |
| P9 | DebtFlag tool to capture, track and manage TD, applicable in scrum sprint planning. | Solution proposal | Tool |
| P10 | Lessons learned from design and implementation of measurement program in agile team where visibility of TD was considered one key issue along with business value, and sprint time estimations. | Experience paper | Lessons learned |
| P12 | Multilevel use of the definition of done is applied to reduce TD in ASD and improve the quality. | Experience paper | Lessons learned |
| P14 | Experience of a cross-functional team using scrum's definition of done to reduce TD. | Experience paper | Lessons learned |
| P15 | Applies work flow (flow based, lean/Kanban development approach) to ensure work items are slowed down enough to have quality (applied analysis with acceptance criteria followed by implementation of code through TDD, code review and testing - exploratory and manual acceptance tests) to reduce TD. | Experience paper | Lessons learned |
| P16 | Systematic agile adoption targeted towards cutting TD (highlights importance of TD assessment plan followed by TD reduction project, assigning special SWAT (special teams that are tasked with evangelizing TD, applying unit testing to reduce TD) | Experience paper | Lessons learned |
| P17 | MAKEFLEXI, to assist architectural flexibility decisions based on TD and real options. It consists of a set of steps to create a model, based on decision trees to estimate when to design for flexibility (decide when to incur TD or strategically payback TD). | Evaluation paper | Model |
| P20 | Method to identify and asses hard to maintain, fault prone code (TD) when developing software code in agile and lean development by measuring code properties (method based on code complexity and revisions of source files) | Evaluation paper | Tool |
| P21 | Describes architecture focused and measurement based approach to strategically manage TD. | Solution proposal | Lessons learned |
| P22 | Investigates TDM strategies in a company adopting agile approaches. | Evaluation | Lessons learned |
| P23 | Shows a case where project solving pattern is seen as a viable option for measuring TD, balancing short and long term values and monitor TD in ASD project. Proposes project solving pattern framework. | Solution proposal | Framework |
| P24 | Identifies TDM as important research area in ASD | Opinion paper | Advice |
| P25 | Proposes combining evolution history information with file dependency structure to detect and locate architecture deviation/degradation, discover shared but undocumented assumptions that cut across module boundaries in ASD. | Evaluation paper | Tool |
| P26 | Suggests adopting service oriented system development to reduce maintenance cost resulting from TD. | Opinion paper | Advice |
| P27 | Suggests 13 steps to minimize TD from developers' perspective (technique implemented to minimize TD) in ASD. | Solution proposal | Guideline |
| P28 | Examined & suggests XP practices to identify & mitigate TD | Experience paper | Lessons learned |
| P30 | Shows experience of XP team in reducing TD by making problems visible using the code Christmas tree. | Experience paper | Lessons learned |
| P32 | Suggests a 2 step approach for managing TD in ASD, i.e. establishing manageable entropy as first step and applying continuous semi-automated quality monitoring and refactoring as 2nd step to support TDM. | Solution proposal | Lessons learned |
| P36 | Introduces the Sqale method to analyze the structure and impact of TD in ASD. | Solution Proposal | Framework |
| P37 | Insight into the organization's strategy for managing and reducing TD in ASD, suggests a technique for visualizing, quantifying and tracking TD (Uses team capacity& velocity to understand the impact of TD; pie and bar charts to visualize, track and understand value of TD). | Solution proposal | Tool |
| P38 | Experience of architecture team in supporting TDM, shows how to visualize high level TD and raise awareness of TD by developers. Proposes the use of TD board to manage and visualize high level TD. | Experience paper | Lessons learned |